\documentstyle[12pt]{article}

\begin{document}
\begin{titlepage}

\begin{center}
{\Large \bf Adaptive Step Size for Hybrid Monte Carlo Algorithm}

\vspace{1cm}
{Philippe de Forcrand {\it and} Tetsuya Takaishi}

{\small \it Swiss Center for Scientific Computing (SCSC) \\
\small \it  ETH-Z\"urich, CH-8092 Z\"urich \\
\small \it  Switzerland \\
}
\end{center}
\vspace{2cm}

\abstract{We implement an adaptive step size method for the Hybrid Monte Carlo algorithm.
The adaptive step size is given by solving a symmetric error equation. An integrator with 
such an adaptive step size is reversible. Although we observe appreciable
variations of the step size, the overhead of the method exceeds its benefits.
We propose an explanation for this phenomenon.
}

\end{titlepage}

\section{Introduction}

Simulations including dynamical fermions remain the most challenging ones 
for lattice QCD.
The standard method to simulate dynamical fermions is, at the moment, the Hybrid Monte Carlo (HMC) 
algorithm\cite{HMC} although it still requires large amounts of computational time.
An alternative method to simulate the dynamical fermions is a local multiboson algorithm based on a polynomial 
approximation of the fermion matrix, proposed by L\"uscher\cite{Luscher}. 
Much interest has been recently devoted to this algorithm\cite{BOSON} to make
it as efficient as the HMC algorithm.

The HMC simulation combines Molecular Dynamics (MD) evolution with a Metropolis
test.
In order to obtain the correct equilibrium, the integrator used in the MD evolution
must satisfy  two conditions; it must be:
\begin{itemize}
\item time-reversible
\item area preserving
\end{itemize}

One such integrator satisfying these conditions is the leapfrog integrator, 
which is normally used in the HMC simulation.
Errors of the leapfrog integrator  start with $O(\Delta t^3)$, where $\Delta t$ is 
the step size of the integrator. These errors cause violation of the conservation of the total energy,
which must be corrected by the Metropolis test at the end of the MD trajectory.
Let $\Delta H$ be the energy violation:
\begin{equation}
\Delta H = H_{end} - H_{begin}
\end{equation}
where $H_{begin}(H_{end})$  is the total enery at the beginning(end) of the MD trajectory.
The Metropolis test accepts a new configuration with a probability $P_{prob}$:
\begin{equation}
P_{prob} \propto \min(1,\exp(-\Delta H)).
\end{equation}
In order to maximize acceptance of the Metropolis test, it might be preferable to 
use a higher order integrator\cite{Creutz}.  
However higher order integrators do not appear practical for lattice QCD
since they require more arithmetic operations 
(force evaluations coming from the fermionic action) than the simplest low-order integrator,
and this overhead exceeds the gain in step-size.

So far conventional HMC simulations have been performed with a fixed step size $\Delta t$ 
during the MD simulation. The local integration error does not remain constant
in this case.
When the trajectory approaches an energy barrier ($S_{eff} = - Log det D$ large),
it is repelled and bounces off. The curvature of the trajectory increases, and
with it the integration error. This situation becomes more pronounced at
small quark mass, since the height of the energy barriers diverges in the
presence of zero modes. Therefore we expect a behavior of the MD trajectory
similar to Fig.1.
Varying the step size adaptively, keeping the local error constant,  
may be a good way to obtain a better integrated trajectory and may result in higher acceptance.
Naively it would seem that this can be accomplished by calculating a local 
error at $(p, U)$, where  
$p=(p_1,p_2,...)$ and $U=(U_1,U_2,...)$ collectively represent momenta and link variables respectively, 
and then by keeping this local error constant.
However this naive scheme is not applicable for the HMC simulation because
it violates reversibility.

Recently an adaptive step size method with reversible structure was proposed 
by Stoffer\cite{Stoffer}. 
He constructed a symmetric error estimator which gives the same error value at a reflected point.
The step size is then determined at every integration point by demanding that the symmetric error estimator 
remain constant.
Stoffer implemented his method for the Kepler problem and obtained better results than the conventional ones.
The possibility to apply this adaptive step size method to the HMC algorithm was stated in Ref\cite{FORC}.
In this letter, we implement this  method for the HMC simulation and 
examine its cost and its efficiency. 

\section{Construction of Adaptive Step Size}

 Here we construct an adaptive step size compatible with a time-reversible
integrator.
 We follow the idea proposed by Stoffer\cite{Stoffer}.

Let $H$ be the Hamiltonian of our system,
\begin{equation}
H= \frac{1}{2}\sum p_i^2 + S(U).
\end{equation}\label{Hamilton}
where $p_i$ are momenta, $U$ are gauge links, 
and $S(U)$ consists of a gauge part $S_g(U)$ and a fermionic part $S_f(U)$,
\begin{equation}
S(U)=S_g(U) + S_f(U),
\end{equation}
\begin{equation}
S_g(U)= \beta \sum (1 - \frac{1}{N_c} ReTr U_{plaq}),
\end{equation}
\begin{equation}
S_f(U)= \phi^\dagger (DD^\dagger)^{-1}\phi,
\end{equation}
where $N_c$ is the number of colors, $\phi$ is a pseudo-fermion vector 
and $D = 1 + \kappa M$ is the 
Wilson fermion matrix, with $\kappa$ the hopping parameter.

 Call $T(\Delta t)$  a one-step integrator.
 It maps momenta and link variables $(p,U)$ onto $(p^\prime,U^\prime)$,
\begin{equation}
T(\Delta t):(p,U) \longrightarrow (p^\prime,U^\prime).
\end{equation}
If this one-step integrator is reversible, then it satisfies
\begin{equation}
T(-\Delta t):(p^\prime,U^\prime) \longrightarrow (p,U),
\end{equation}
In this study, we  use  the leapfrog integrator as our one-step integrator.
In terms of the time evolution operators\cite{Creutz,SW}, 
the one-step integrator $T(\Delta t)$ can be written as 
\begin{equation}
 T(\Delta t) = exp( \frac{\Delta t}{2} L(\frac{1}{2}\sum p_i^2)) 
exp(\Delta t L(S(U)))  exp( \frac{\Delta t}{2} L(\frac{1}{2}\sum p_i^2)),
\label{LEAP}
\end{equation}
where $L(\bullet)$ is the linear operator which is given by the Poisson bracket\cite{SW}.
\sloppy{The one-step integrator requires one force evaluation represented by 
$exp(\Delta t L(S(U)))$.} The fermionic part of the force depends on the
solution of a linear equation of the type $D x = \phi$, which is obtained
iteratively at great expense of computer time.
Thus force evaluations dominate the computation, and the cost of our
algorithm can be measured in units of force evaluation.

 Now we define a symmetric error estimator,
\begin{equation}
E_S (p,U:\Delta t) = e(p,U:\Delta t) + e(p^\prime, U^\prime : -\Delta t),
\label{ES}
\end{equation} 
where $e(p,U:\Delta t)$ is a local error at $(p,U)$ when the system is integrated by some integrator
with a step size $\Delta t$, and  the integrator maps $(p,U)$ on $(p^\prime, U^\prime)$.
The local error is assumed to increase monotonically with $\Delta t$.
We will define the local error later.
If the integrator is reversible, eq.(\ref{ES}) is obviously symmetric under the exchange:
\begin{equation}
(p,U,\Delta t) \longleftrightarrow (p^\prime, U^\prime,-\Delta t).
\end{equation}
Namely, this means 
\begin{equation}
E_S(p,U:\Delta t) = E_S(p^\prime, U^\prime:-\Delta t).
\end{equation}
The adaptive step size is then determined by solving a symmetric error equation,
\begin{equation}
\label{tol}
E_S(p,U:\Delta t) = tolerance
\end{equation}
The tolerance should be kept constant during the MD simulation.
The adaptive step size determined by eq.(\ref{tol}) takes 
the same value at the reflected point $(-p^\prime, U^\prime)$.
Therefore we find that an integrator with the adaptive step size determined by eq.(\ref{tol}) is reversible.

Any local error can be defined provided that it increases monotonically with $\Delta t$.
Our local error is defined as follows.

First, we integrate $(p,U)$ by the two-step integrator $T^2(\Delta t)$ and 
the one-step integrator $T(2\Delta t)$.
\begin{equation}
T^2(\Delta t):(p,U) \longrightarrow  (p^\prime, U^\prime)
\label{def1}
\end{equation}
\begin{equation}
T(2\Delta t):(p,U) \longrightarrow  (\tilde{p^\prime}, \tilde{U^\prime})
\end{equation}
If $\Delta t$ is not too large, $(p^\prime, U^\prime)$ and $(\tilde{p^\prime}, \tilde{U^\prime})$
should be close to each other. 
We define the local error at $(p,U)$ by
\begin{equation}
e(p,U:\Delta t) = \sum_{\mu, x}^{4, V}(1-\frac{1}{N_c}Re Tr {U^\prime}^\dagger_\mu(x) 
\tilde{U^\prime_\mu}(x))/(4 V)
\label{local}
\end{equation}
where $V$ is the volume of the lattice.
One could also use the momenta in the definition of the local error.
Similarly, we integrate $(p^\prime, U^\prime)$ by $T^2(-\Delta t)$ and $T(-2\Delta t)$
in the inverse time direction,
\begin{equation}
T^2(-\Delta t):(p^\prime, U^\prime) \longrightarrow  (p, U)
\label{T2}
\end{equation}
\begin{equation}
T(-2\Delta t):(p^\prime, U^\prime) \longrightarrow  (\tilde{p}, \tilde{U}).
\end{equation}
Since the integrator is reversible, the calculation of eq.(\ref{T2}) is not needed.
The local error at $(p^\prime, U^\prime)$ is also defined like eq.(\ref{local}),
\begin{equation}
e(p^\prime,U^\prime:-\Delta t) = \sum_{\mu, x}^{4, V}(1-\frac{1}{N_c}Re Tr {U}^\dagger_\mu(x)
\tilde{U_\mu}(x))/(4 V)
\label{def2}
\end{equation}
In the case of the leapfrog integrator of eq.(\ref{LEAP}), we need 4 force evaluations to 
construct the symmetric error estimator $E_S$, instead of just 2 for the
evolution $T^2(\Delta t)$.

Eq.(\ref{tol}) is a non-linear equation.
In general, it should be solved numerically, eg. by iterative bisection.
With our definition of the symmetric error estimator, however, we can 
anticipate the scaling behavior of eq.(\ref{tol}) and use it to accelerate convergence.
The vector potentials evolved by the leapfrog integrator have $O(\Delta t^3)$ errors,
\begin{equation}
\tilde{A}_\alpha^\prime = A_\alpha^\prime +  O(\Delta t^3). 
\end{equation}
Therefore,
\begin{eqnarray}
{U^\prime}^\dagger \tilde{U^\prime} & = & exp( -i A^\prime_\alpha \lambda_\alpha)
exp( i \tilde{A^\prime_\alpha} \lambda_\alpha) \\
& \approx & 1 + i c_\alpha \lambda_\alpha O(\Delta t^3) + d_{\alpha\beta} \lambda_\alpha\lambda_\beta 
O(\Delta t^6) 
\label{RT}
\end{eqnarray}
where $\lambda_\alpha$ are  SU(3) generators, and  
$c_\alpha$ and $d_{\alpha\beta}$ are some real constants whose explicit values are not important here.  
Taking $Real$ and $Trace$ of eq.(\ref{RT}) and substituting it to eq.(\ref{local}) and (\ref{def2}), 
we find that in the leading order the symmetric error 
estimator starts with $O(\Delta t^6)$.
This behavior is verified numerically, as illustrated in Fig.2: 
if $\Delta t$ is not too large, the symmetric error estimator behaves like $E_S \propto \Delta t_A^6$.
This property is used for solving eq.(\ref{tol}). Choose some initial
$\Delta {t_A}_1$ for the step-size and calculate ${E_S}_1=E_S(\Delta {t_A}_1)$.
If $\Delta {t_A}_1$ does not satisfy the symmetric error equation eq.(\ref{tol}) 
then input the second trial value $\Delta {t_A}_2$, 
which is the solution of
\begin{equation}
log(\frac{tolerance}{{E_S}_1}) = 6 log(\frac{\Delta {t_A}_2}{\Delta {t_A}_1}).
\end{equation}
If further trials are necessary, the following approximation can be used \cite{Stoffer},
\begin{equation}
log(\frac{\Delta {t_A}_3}{\Delta {t_A}_2})=\frac{log(\Delta {t_A}_2/\Delta {t_A}_1)}{log({E_S}_2/{E_S}_1)}
log(\frac{tolerance}{{E_S}_2}).
\end{equation}
This recurrence is continued until eq.(\ref{tol}) is satisfied to sufficient accuracy,
and then a new configuration  $(p^\prime, U^\prime)$ is stored. 
Note that 2 strategies are available, just like for the stopping criterion
of the linear solver in the force calculation: 
either the initial guess $\Delta {t_A}_1$ 
is invariant under time-reversal (eg. it is equal to the average step-size), 
and the accuracy to which eq.(\ref{tol}) must be satisfied 
can be set arbitrarily low; or the initial guess makes use of past information
(eg. it is equal to the previous step-size), and eq.(\ref{tol}) must be
satisfied exactly. 
We use the second method, and take the previous result as our initial guess.
We then solve eq.(\ref{tol}) to within 5\%. Since we do not solve 
eq.(\ref{tol}) exactly, we introduce a tiny, controllable source of 
irreversibility in the dynamics: the step size under time-reversal could
be different by about $5/6 \sim 1 \%$. For this exploratory study, we have 
not considered this aspect further.

\section{Efficiency}
\sloppy{We examine the method for full QCD ($N_c = 3$) at several parameters 
$(\beta,\kappa,volume)$} listed in Table 1.
We chose $\beta = 0$ in several instances, to eliminate the gauge part of
the action and hopefully be more sensitive to the energy barriers coming
from the fermionic part. The effect of changing the quark mass can be
obtained by comparing cases A and B, that of changing the volume by 
comparing A and D.
The adaptive step size is determined 
by the symmetric error equation eq.(\ref{tol}), 
with the tolerance set as per Table 1. 
The adaptive step sizes are summed up from the beginning of the trajectory and
when the total trajectory length becomes greater than the trajectory length of Table 1, 
a Metropolis test is performed.

Fig.3 shows histograms of the adaptive step size at $\beta=0.0$, $\kappa=0.215$ and 0.230
( cases A and B in Table 1 ).
The distribution remains strongly peaked. This is also true of cases C and D.
The average step size $<\Delta t_A>$ and its relative variance are summarized in Table 2,
where the relative variance $\sigma$ is defined by
\begin{equation}
\sigma^2 = \frac{1}{N}\sum_{i=1}^{N} (\frac{{\Delta t_A}_i}{<\Delta t_A>} - 1)^2.
\end{equation}
As the quark mass $m_q$ decreases, the energy barriers in phase space become
higher and sharper, so that one would expect the variation of the step size
to increase. Indeed this is what happens, and the relative variance in cases
A and B increases roughly like $1/m_q$, where 
$m_q \propto (\kappa^{-1} - \kappa_c^{-1})$ and $\kappa_c = 1/4$ at $\beta = 0$.
On the other hand, as the volume increases, the relative variance seems to
decrease sharply, like $1 / \sqrt V$ or faster (see cases A and D).
Perhaps this can be explained by considering the relative fluctuations of the 
effective action $-Log~det~D$: as the volume increases, the relative fluctuations
decrease, so that the system tends to stay at some average distance from the
energy barriers, rather than bouncing off them.

From the schematic picture of Fig.1, it is expected that the approach of
an energy barrier causes a reduction in the adaptive step size,
and at the same time an increase in the number of iterations 
taken by the solver to converge.
Fig.4 shows $\Delta t_A$ versus the number of iterations in the solver:
the expected anti-correlation between them can be observed, and becomes
more pronounced as the quark mass is reduced.

In order to compare the adaptive method with the conventional HMC algorithm,
we define the efficiency of the adaptive method as follows.

First, find the fixed step-size $\Delta t_{HMC}$ of the HMC simulation 
which gives the same acceptance as the adaptive step-size method.  
The total trajectory length of the HMC simulation is set to the average
total trajectory length $<$traj. length$>$ of the corresponding adaptive step size method's 
case. We performed the HMC simulations with several step-sizes
and determined the corresponding  step-size of the HMC algorithm by 
interpolating those results.
The results of the corresponding  step-size are summarized in Table 2. 
For the acceptance of the adaptive step-size method, see Table 2.

Then, define the gain by
\begin{equation}
g_A = <\Delta t_A> / \Delta t_{HMC}.
\end{equation}
When $g_A > 1$, the adaptive step-size method really takes larger steps on
average, without compromising the acceptance.
However the real efficiency of the method can only be assessed by taking into
account the overhead of determining the adaptive step-size, since
additional force evaluations are necessary.
  
From the definition of our symmetric error estimator 
eq.(\ref{def1})-(\ref{def2}), we know that one construction of $E_S$ 
needs 4 force evaluations. Call $R_T$ the average number of trial steps 
needed to solve eq.(\ref{tol}). After $4 R_T$ force evaluations, we
have determined the step-size $\Delta t_A$ and use the integrator 
$T^2(\Delta t_A)$ (eq.\ref{def1}) to advance the dynamics by 2 steps $\Delta t_A$.
Therefore the cost of the method is $2 R_T$ force evaluations per step,
compared with $1$ force evaluation per step for standard HMC.

Real efficiency will be achieved if the number of force evaluations per
unit time decreases, ie. if $2 R_T < g_A$.
Results for the gain $g_A$ and the cost $2 R_T$ are summarized in Table 3.
For all cases we studied, real efficiency is not achieved.

It is disappointing to see how small the gains $g_A$ are.
The reason for such small gains can be understood by considering   
the behavior of the Hamiltonian.
The dependence of the energy violation at each step with the step-size is, 
in general, non-linear.
Therefore it is not necessary that a small local error correspond to 
a proportionally small energy violation of the Hamiltonian.
Fig.5 shows $|\Delta H|$, the absolute value of the energy violation after 
one integration step, versus the local error $E_S$.
The 2 clusters of points correspond to fixed step sizes $\Delta t=0.04$ 
and $0.08$.
No strong correlation between $E_S$ and $\Delta H$ can be observed.
This is further evidenced by the dashed lines, which are the result of
fitting  to a scaling law  $\sqrt{<\Delta H^2>} = E_S^b$:
for the larger step-size, $b$ is almost zero.
Therefore, it becomes clear that fixing $E_S$ and varying $\Delta t$ adaptively 
cannot have a strong effect on the acceptance, 
which solely depends on $\Delta H$.

Two approaches could be used to improve the efficiency of our scheme:\\
i) decrease the overhead: instead of estimating the error by comparing
$T^2(\Delta t)$ with $T(2 \Delta t)$, one could replace the latter by
an Euler integrator, which requires no additional force evaluation. Note
that the error eq.(\ref{ES}) remains symmetric under the exchange
$(p,U,\Delta t) \longleftrightarrow (p^\prime, U^\prime,-\Delta t)$
even though the Euler integrator is not time-reversible. 
The problem we found with that
approach is that, for the large step-sizes used on our small lattices,
the error eq.(\ref{ES}) no longer obeyed a simple scaling law eq.(\ref{RT}) as
a function of the step-size. Then the number of iterations needed to
solve eq.(\ref{tol}) increased, defeating the expected reduction in overhead.
On larger lattices with smaller step-sizes, this problem would be milder.\\
ii) change the definition of the error eq.(\ref{def2}), so that it is better
correlated with $|\Delta H|$, the energy violation at each step. 
Note that $|\Delta H|$ itself cannot be 
chosen, because it does not increase monotonically with the step-size:
in that case eq.(\ref{tol}) admits multiple solutions; the overhead of
converging to one of them, and the same one under time-reversal, increases
considerably.
With our definition eq.(\ref{def2}), the error is only weakly correlated 
with $|\Delta H|$, but the situation again
seems to improve with smaller step-sizes, on larger lattices
(compare the 2 dashed lines in Fig.5). Nonetheless it would be desirable
to control the step-size with a more relevant quantity than eq.(\ref{def2}), 
since all that matters in the end is energy conservation. 

Finally, instead of varying the step-size, one could vary adaptively
the couplings of the Hamiltonian $H$, eq.(\ref{Hamilton}), at each step, or even
include some new operators in $H$, trying to tune them so as to best conserve
energy. The general difficulty with that approach is to find an error
eq.(\ref{def2}) which varies monotonically with the couplings of $H$.

\section{Conclusion}
We have implemented an adaptive step-size method for Hybrid Monte Carlo 
simulations, and tested it at several parameters $(\beta,\kappa, volume)$.
The relative variance of the step-size increases for small quark masses and
small volumes. The average step-size seems somewhat larger than the
corresponding fixed step-size at same acceptance. But this gain is more
than offset by the overhead of determining the adaptive step-size.
It seems very difficult to achieve real gains in efficiency, because 
conservation of energy, which is necessary for high Metropolis acceptance
in the HMC algorithm, is poorly correlated with the conventional error 
governing the adaptive step-size.

A plausible extrapolation from our results would indicate that the 
relative variance of the step-size scales like $m_q^{-1} V^{-1/2}$, ie.
as $(m_{\pi} L)^{-2}$, where $m_{\pi}$ is the pion mass and $L$ the physical
size of the lattice. This quantity normally remains constant as the continuum
limit $a \rightarrow 0$ of the lattice theory is taken, so that the relative
fluctuations in the adaptive step-size $\Delta t$ would tend to a constant. 
Even if this analysis is no more than plausible at this stage, it is clear 
that the two limits $m_q \rightarrow 0$, $V \rightarrow \infty$ have opposite 
effects on the fluctuations of $\Delta t$, making it unlikely that such 
fluctuations become very large on present lattice sizes. This observation is 
consistent with the limited fluctuations (a factor 2 or so) in the number of 
iterations needed by the solver to compute the force in the largest QCD
simulations \cite{SESAM}.

Thus it appears that QCD is much ``easier'' to simulate than the Kepler
problem: in lattice QCD, the force on the gauge links varies little in 
magnitude, and the curvature of the Molecular Dynamics trajectory is rather
small. One intuitive explanation is that the QCD force is dominated by
short-range UV contributions, which drown the IR component sensitive to
the energy barrier $det D \sim 0$.

\vspace{1cm}
We thank D. Stoffer for helpful discussions.
T.T. is supported in part by the Japan Society for the Promotion of Science.
Ph. de F. thanks Hiroshima Univ. and Tsukuba Univ., especially 
Profs. O. Miyamura and Y. Iwasaki, for hospitality during this project.

\newpage
{\Large \bf Figure Captions}

Figure 1: Schematic behavior of a Molecular Dynamics trajectory in 
configuration space. $detD$ is the determinant of the Dirac matrix.

Figure 2: Adaptive step size $\Delta t_A$ versus symmetric error $E_S$, for
two configurations of size $4^4$ at $\kappa=0.230$.
The straight line $E_S \propto \Delta t_A^6$ is shown for comparison.

Figure 3: Histograms of adaptive step size $\Delta t_A$, for $4^4$ lattices at 
$\kappa=0.215$ and 0.230. The logarithmic scale shows the increase with
$\kappa$ of the relative variance.

Figure 4: Adaptive step size versus number of iterations in the solver (BiCG$\gamma_5$).

Figure 5: $|\Delta H|$ versus the local error $E_S$, on a $4^4$ lattice at $\kappa=0.215$. 
The step size is fixed at $\Delta t=0.04$ and $0.08$. 
$\Delta H$ is the change in the total energy after one integration step.
The dotted lines result from fitting to
the form $\sqrt{<\Delta H^2>} = E_S^b$, and show the correlation (or absence of)
between the 2 quantities. $<\Delta H^2>$ is obtained by dividing the data 
in 10 bins and averaging the values $\Delta H^2$ in each bin.

\newpage
\begin{table}
\begin{tabular}{|c|ccccc|} \hline
case & $\beta$ & size  & $\kappa$ & traj. length & tolerance($\pm5\%$) \\ \hline
A    & 0.0     & $4^4$ & 0.215    &  0.8         & $10^{-4}$              \\ 
B    & 0.0     & $4^4$ & 0.230    &  0.4         & $8\times 10^{-6}$      \\ 
C    & 5.4     & $4^4$ & 0.162    &  1.0         & $10^{-6}$              \\ 
D    & 0.0     & $8^4$ & 0.215    &  0.3         & $10^{-7}$              \\ \hline
\end{tabular}
\caption{Run parameters}
\end{table}

\begin{table}
\begin{tabular}{|c|ccccc|} \hline
case & $<\Delta t_A>$ & $\sigma(\%)$  & $<$traj. length$>$ & acceptance(\%) & $\Delta t_{HMC}$ \\ \hline
A    & 0.0911(3)     & 3.3      & 0.91 & 36(2) & 0.0897(13)         \\
B    & 0.0431(1)     & 5.6      & 0.44 & 57(2) & 0.0419(09)         \\
C    & 0.0673(1)     & 0.8      & 1.08 & 87(2) & 0.0688(80)         \\
D    & 0.03281(2)    & 0.6      & 0.33 & 63(2) & 0.0328(10)         \\ \hline
\end{tabular}
\caption{Results of the adaptive step-size method and fixed step-size $\Delta t_{HMC}$ of the
HMC algorithm.  
$<$traj. length$>$ stands for the average total trajectory length.
The fixed step-size $\Delta t_{HMC}$ is determined so that it gives 
the same acceptance as the adaptive step size method. 
}
\end{table}

\begin{table}
\begin{tabular}{|c|ccc|} \hline
case & $g_A$  & $R_T$  & Cost per step $(=2R_T)$  \\ \hline
A    & 1.016(15)     & 2.25   & 4.50    \\
B    & 1.029(22)     & 2.45   & 4.90    \\
C    & 1.0(1)     & 1.13   & 2.26    \\
D    & 1.00(3)     & 1.15   & 2.30    \\ \hline
\end{tabular}
\caption{Gain $g_A$, average number of trial steps $R_T$ and Cost per step}
\end{table}

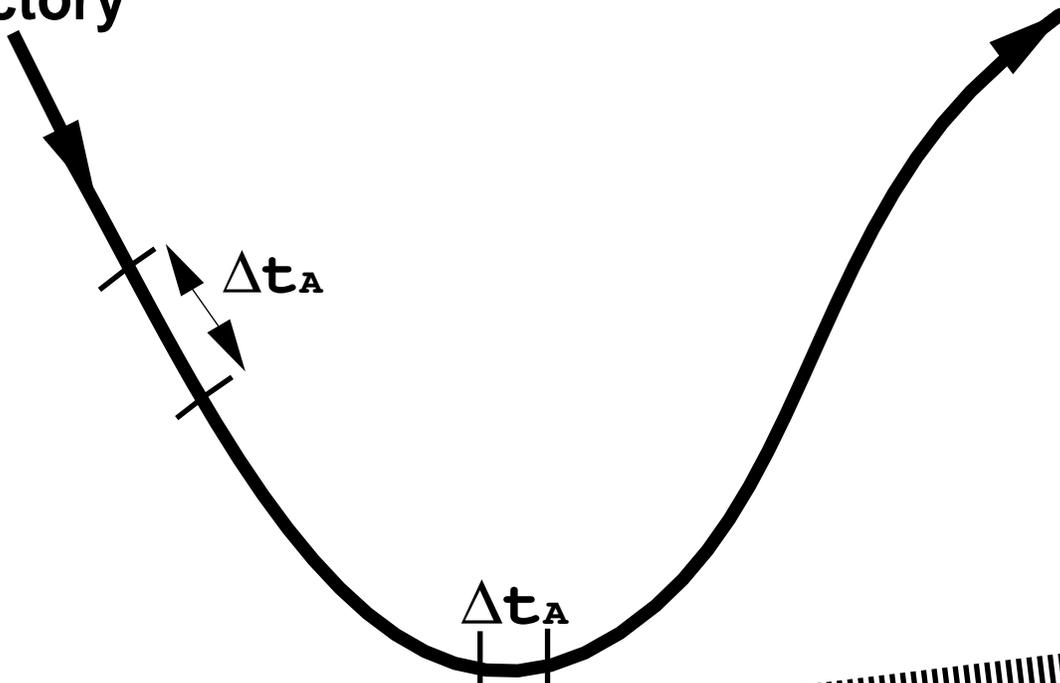
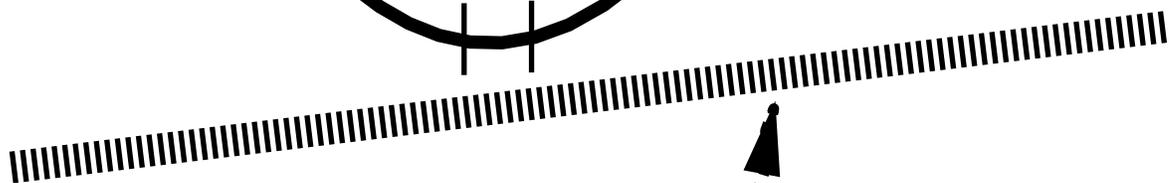
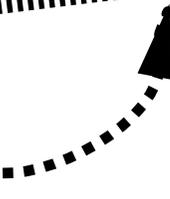
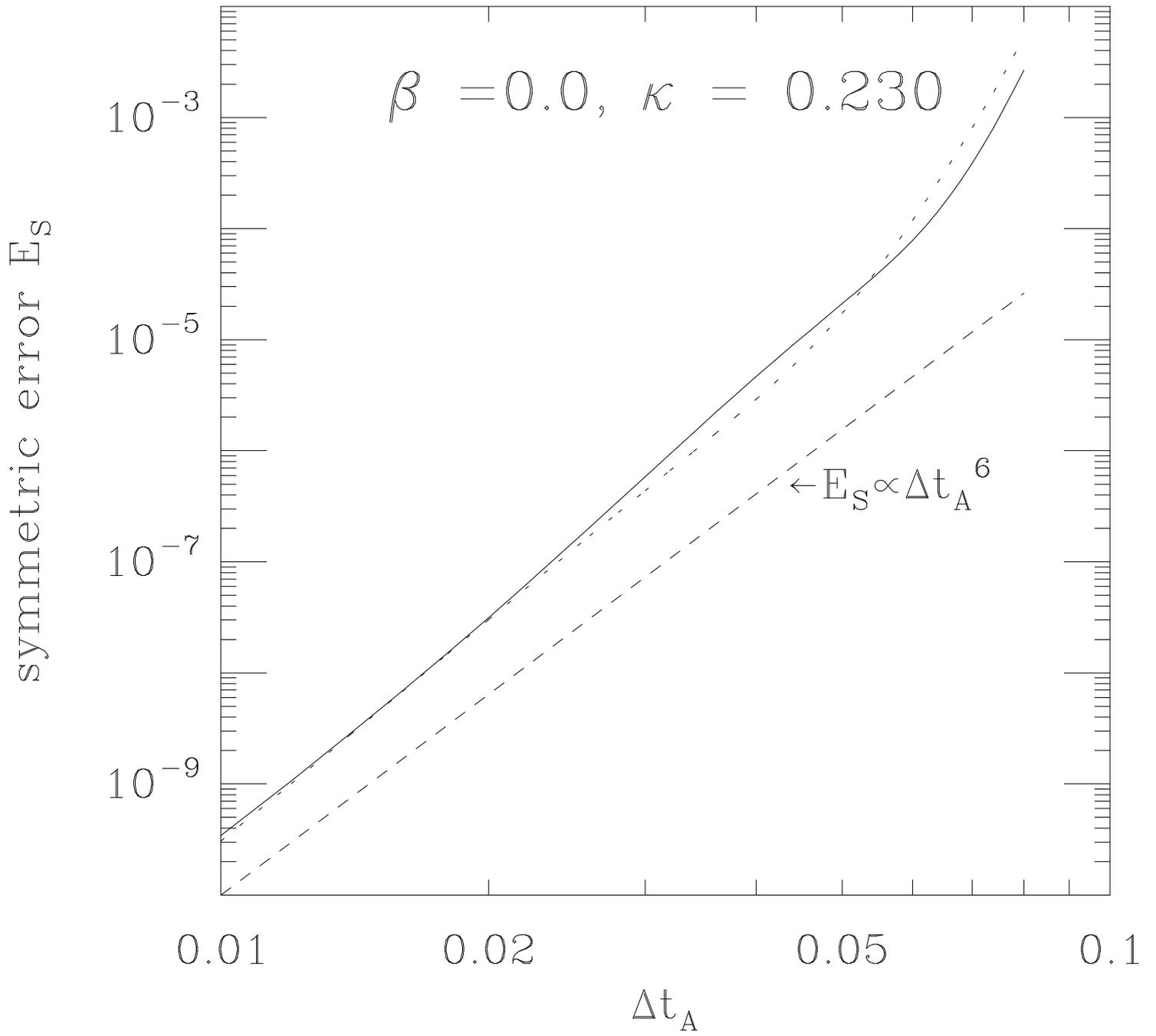
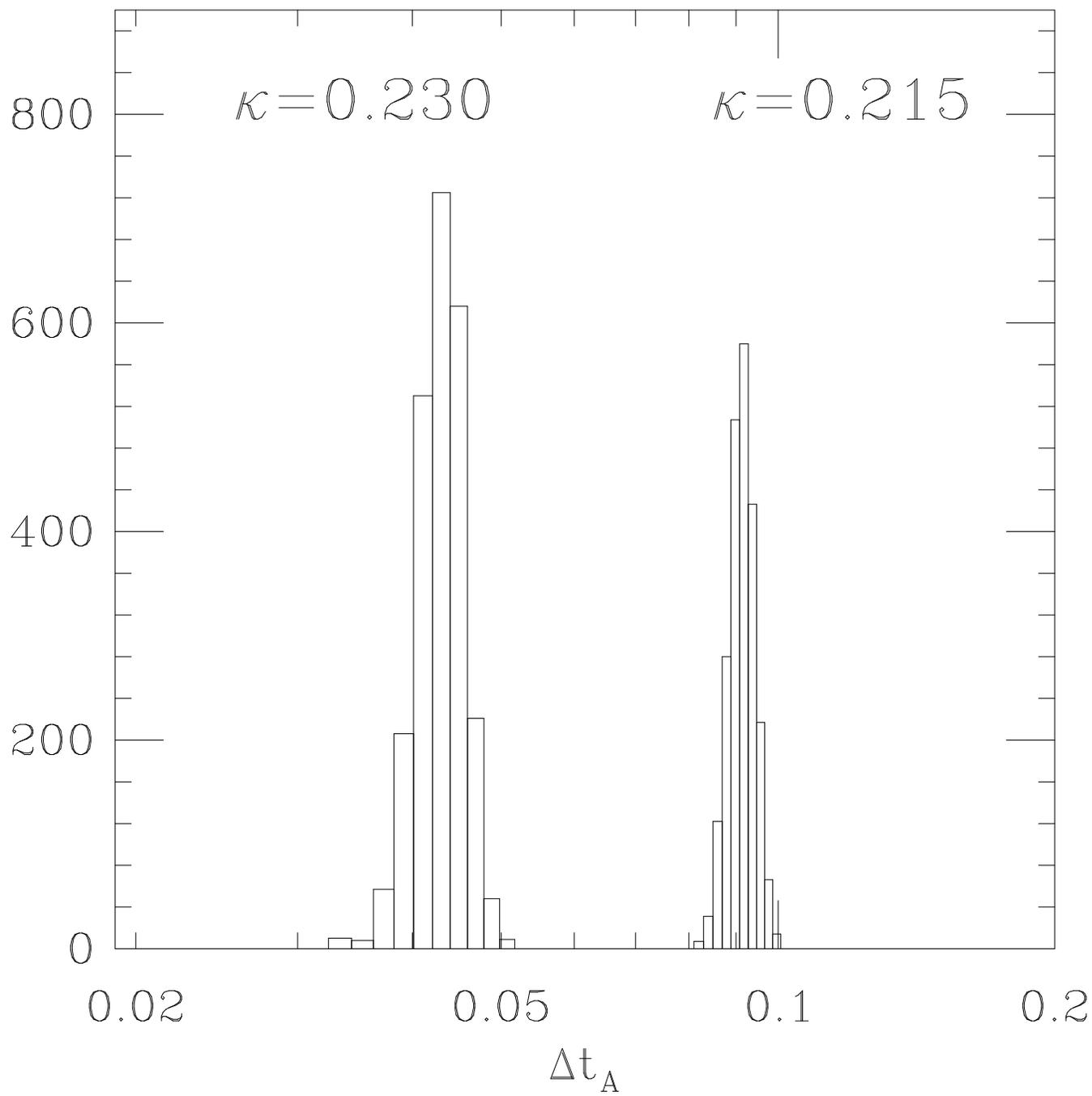
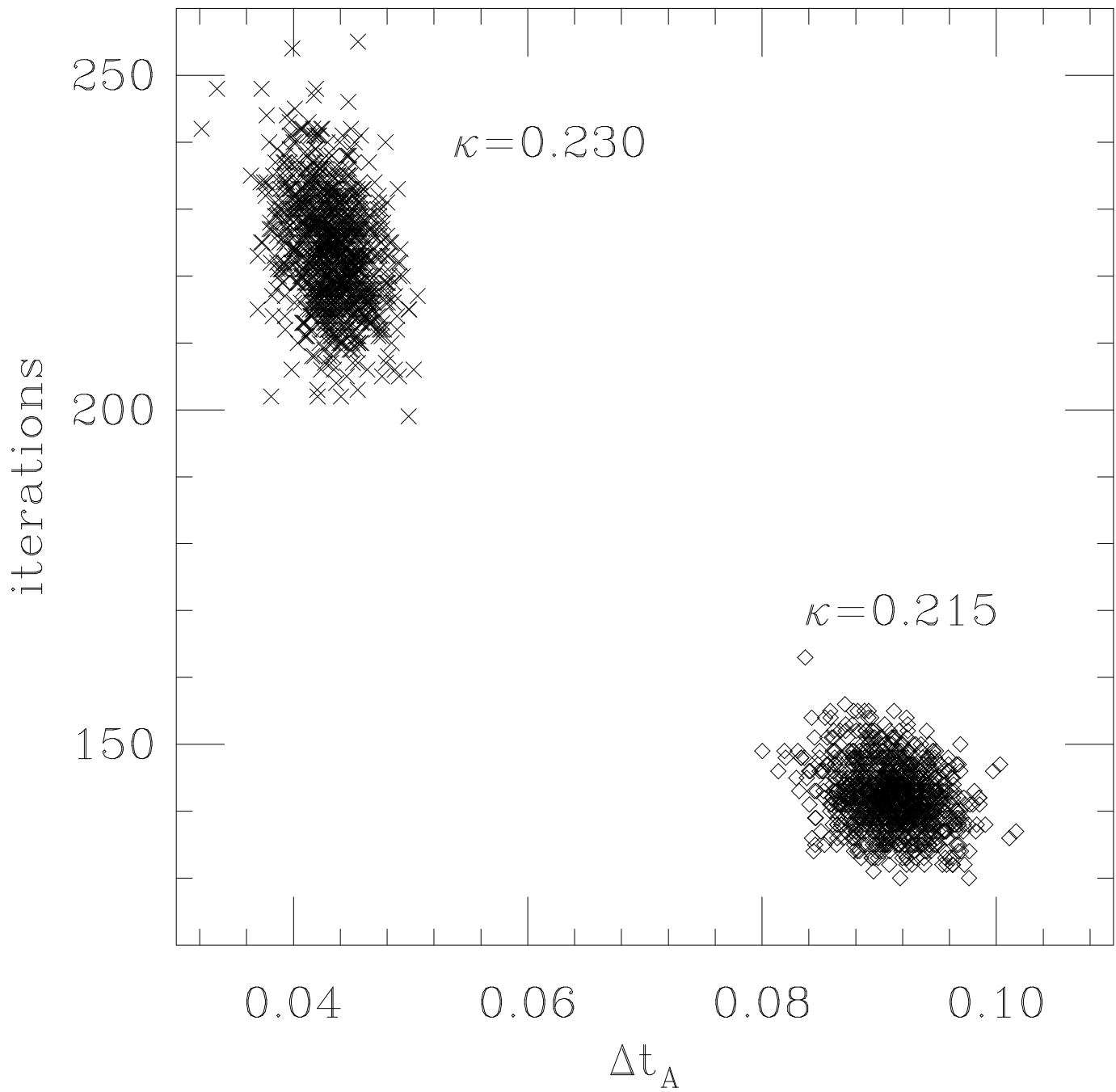
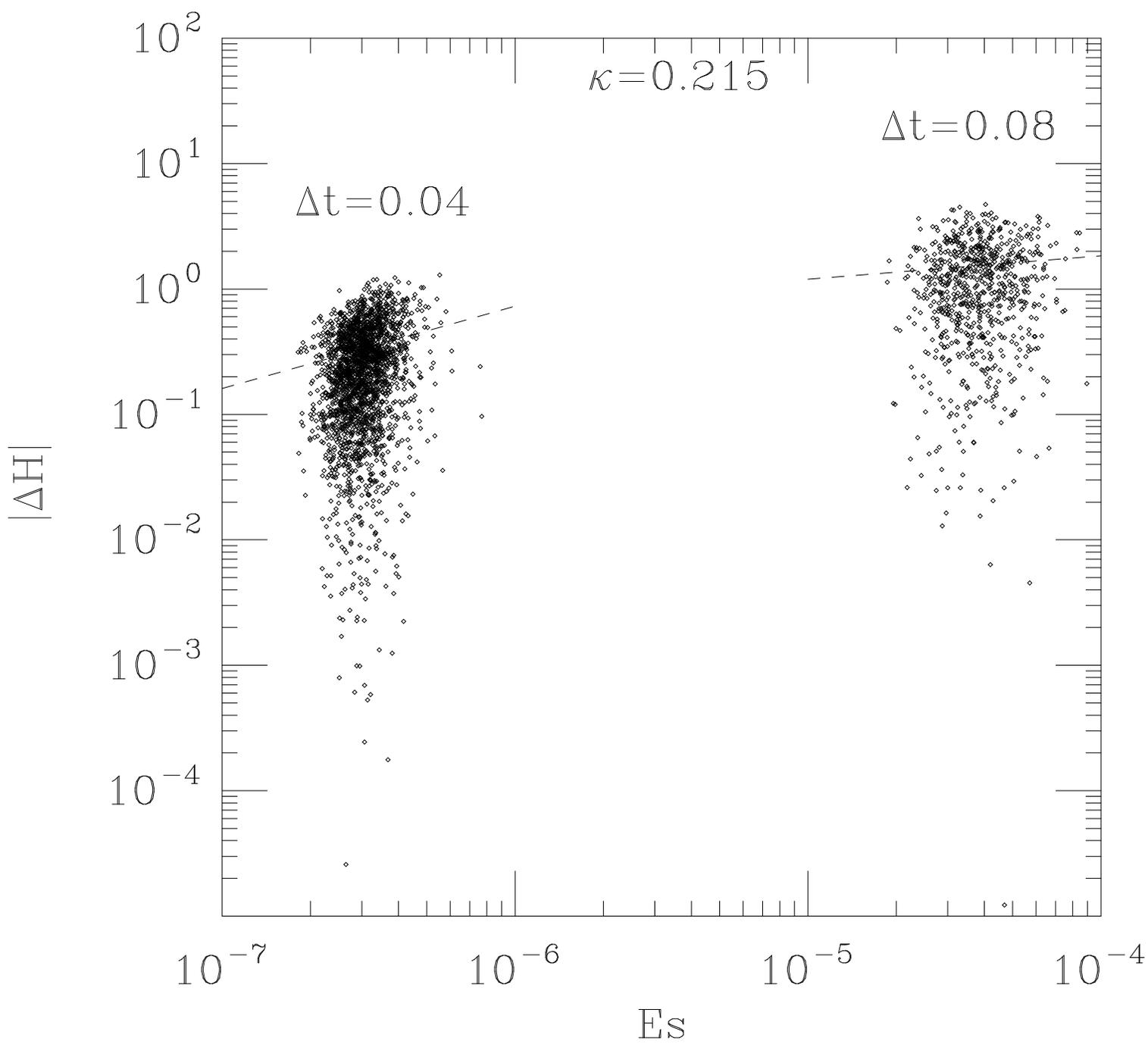
\end{document}